\documentclass[10pt,twocolumn]{article}

\usepackage{geometry}
\usepackage{amsmath,amssymb,amsfonts}
\usepackage{bm}

\usepackage{graphicx}
\usepackage{subfigure}
\usepackage{booktabs}
\usepackage{multirow}
\usepackage{array}
\usepackage{makecell}
\usepackage{float}

\usepackage{tikz}
\usetikzlibrary{shapes,arrows,positioning,fit,backgrounds,calc,decorations.pathreplacing,patterns}
\usepackage{pgfplots}
\pgfplotsset{compat=1.17}

\usepackage{algorithm}
\usepackage{algorithmic}

\usepackage{listings}
\usepackage{xcolor}
\usepackage{hyperref}
\hypersetup{colorlinks=true,linkcolor=black,citecolor=black,urlcolor=blue}

\usepackage{enumitem}

\usepackage{amsthm}
\newtheorem{theorem}{Theorem}

\newtheorem{property}{Property}
\newtheorem{lemma}{Lemma}

\begin{document}

\title{HCMS: Head-Chunked Multi-Stream Pipeline for Communication-Computation Overlap in Long-Sequence Parallel Attention}

\author{
\textbf{Chao Yuan}$^{1}$, \textbf{Pan Li}$^{1}$, \textbf{Yingnan Sun}$^{1}$, \textbf{Jing Liu}$^{1}$ \\[0.5em]
$^{1}$Bilibili Inc., Shanghai, China \\
{\small \{yuanchao, lipan02, sunyingnan, liujing04\}@bilibili.com}
}

\date{}

\maketitle

% ============================================================================
% Abstract
% ============================================================================
\begin{abstract}
All-to-all based sequence parallelism methods execute communication and computation strictly in serial when processing medium-long sequences, resulting in hardware resource underutilization. This paper proposes Head-Chunked Multi-Stream Pipeline (HCMS), which exploits the computational independence of multi-head attention by partitioning attention heads into multiple chunks and achieving fine-grained communication-computation overlap through dual CUDA streams. HCMS is orthogonally compatible with existing optimizations such as FlashAttention and SDPA, requires no modification to underlying kernels, supports uneven partitioning while maintaining numerical equivalence. Experiments validate the effectiveness across four GPU platforms at 2-8 GPU scales: for typical video generation sequence lengths of 31K-56K tokens, HCMS achieves 10\%-17.5\% speedup over the Ulysses baseline and 5\%-14.5\% speedup over Ring Attention; end-to-end acceleration of 6.8\% is achieved on the Wan2.2 model. Theoretical analysis shows that HCMS benefits are positively correlated with communication ratio $\rho$, and its use is recommended when $\rho>20\%$.
\end{abstract}

\vspace{0.3em}

% ============================================================================
% 1. Introduction
% ============================================================================
\section{Introduction}

Large-scale Transformer models~\cite{vaswani2017attention} have achieved breakthrough advances in natural language processing~\cite{devlin2019bert,brown2020language}, computer vision~\cite{dosovitskiy2021image}, and multimodal generation. Video generation represents one of the most prominent application scenarios~\cite{hong2022cogvideolargescalepretrainingtexttovideo,blattmann2023stable,singer2022make,ho2022imagen}. Taking models such as Sora~\cite{openai2024sora} and Wan2.2 as examples, generating 2-4 second videos at 720P resolution corresponds to sequence lengths of approximately 31K-56K tokens in latent space. This medium-long sequence processing requirement makes sequence parallelism a critical technique---distributing sequences across multiple GPUs to accelerate computation.

However, all-to-all based sequence parallelism schemes, represented by DeepSpeed Ulysses~\cite{jacobs2023deepspeedulyssesoptimizationsenabling}, exhibit significant efficiency issues: communication and computation execute strictly in serial, causing hardware resource underutilization. We observe that under typical configurations of 4-8 GPUs with PCIe interconnect and 31K-100K token sequence lengths, the communication ratio $\rho$ typically ranges from 15\% to 40\%. This characteristic provides substantial room for communication optimization---through communication-computation overlap, a theoretical speedup upper bound of $1/(1-\rho)$ can be achieved.

Existing sequence parallelism methods each have their limitations. Ring Attention~\cite{liu2023ringattentionblockwisetransformers} employs a ring communication pattern where overlap depends on block-level pipelined execution; when the number of blocks is small, overlap effectiveness is limited, and $P-1$ rounds of serial communication are required. Although DeepSpeed Ulysses requires fewer communication rounds, its original implementation executes communication and computation completely in serial, failing to exploit overlap optimization.

A fundamental property of multi-head attention is that computations across different heads are mutually independent. This independence implies that the serial dependency of ``communication $\rightarrow$ computation'' can be relaxed at the head granularity---rather than waiting for all heads' input data to complete communication before starting computation, computation for any head can begin immediately once its data is ready. This property provides the theoretical foundation for achieving fine-grained communication-computation overlap at the head dimension.

Based on this insight, we propose the Head-Chunked Multi-Stream Pipeline (HCMS) method, which partitions attention heads into multiple independent chunks and achieves fine-grained pipelining through dual CUDA streams. HCMS supports uneven chunk partitioning with performance variance less than 1\%, and is orthogonally compatible with existing attention optimizations such as FlashAttention and SDPA without requiring kernel modifications. Experiments demonstrate that HCMS achieves significant speedups across four GPU platforms at 2-8 GPU scales: for typical video generation sequence lengths of 31K-56K tokens, 4-GPU configurations achieve 17.5\% speedup and 8-GPU configurations achieve 16.4\% speedup, outperforming Ring Attention by 5\%-14.5\%. In end-to-end validation on the Wan2.2 video generation model, attention layers achieve 18.3\% speedup and the complete pipeline achieves 6.8\% speedup.

The main contributions of this paper are as follows:
\begin{itemize}
    \item \textbf{Head Chunking Partitioning Strategy:}
    We leverage the computational independence of multi-head attention to partition attention heads into $C$ chunks, thereby decoupling the originally serial communication-computation dependency into $C$ independently schedulable subtasks, achieving fine-grained parallelism without altering computational semantics. This method supports uneven partitioning (when $H \bmod C \neq 0$), and experiments show that performance variance compared to even partitioning is less than $1\%$.

    \item \textbf{Theoretical Modeling and Applicability Analysis:}
    We establish a performance model for HCMS and derive the optimal chunk count $C^* = \sqrt{T_{comm}/\beta}$, clearly defining its applicability: significant benefits can be achieved when communication ratio $\rho > 20\%$, while benefits are limited when $\rho < 10\%$.

    \item \textbf{Cross-Platform Experimental Validation:}
    Across four GPU platforms with $2$--$8$ GPU configurations, for sequence lengths of $31\text{K}$--$56\text{K}$ tokens, HCMS achieves $5\%$--$14.5\%$ speedup over Ring Attention and $10\%$--$17.5\%$ speedup over the Ulysses baseline. Furthermore, the method is fully compatible with PyTorch autograd and can be directly used for distributed training, achieving $3.5\%$ training speedup while maintaining numerical equivalence.
\end{itemize}

% ============================================================================
% 2. Related Work
% ============================================================================
\section{Related Work}

\subsection{Sequence Parallelism Methods}

Sequence parallelism distributes the sequence dimension across multiple devices~\cite{li2021sequence,li2023lightseq}, primarily including ring communication-based methods and all-to-all based methods. The computations of different heads in multi-head attention are mutually independent, and this computational independence forms the theoretical foundation of our method.

Ring Attention~\cite{liu2023ringattentionblockwisetransformers} employs a ring communication pattern, partitioning sequences into $P$ blocks and achieving distributed attention computation through $P-1$ rounds of KV block passing. Its advantage lies in high memory efficiency, enabling processing of arbitrarily long sequences. However, Ring Attention's communication-computation overlap is limited by block granularity: overlap is only effective when computation time exceeds communication time. Furthermore, $P-1$ rounds of serial communication make total latency proportional to $P$. Subsequent work such as Striped Attention~\cite{liu2023striped} optimized for causal attention, but the video generation scenarios we focus on use bidirectional attention, making these optimizations not directly applicable.

DeepSpeed Ulysses~\cite{jacobs2023deepspeedulyssesoptimizationsenabling} employs an all-to-all communication pattern: the input all-to-all rearranges data from ``sequence-sharded, heads-complete'' to ``sequence-complete, heads-sharded'', and after attention computation, the output all-to-all performs the inverse transformation. Compared to Ring Attention's $P-1$ communication rounds, Ulysses requires only 2 rounds of all-to-all, with fewer communication rounds and better compatibility with FlashAttention~\cite{dao2022flashattention,dao2024flashattention2}. However, in Ulysses' original implementation, communication and computation execute completely in serial, which is precisely the optimization target of this paper.

Megatron Context Parallelism~\cite{shoeybi2019megatron,narayanan2021efficient} provides both ring communication and all-gather modes. USP~\cite{fang2024uspunifiedsequenceparallelism} unifies Ring and Ulysses into a single framework and proposes tile-based communication overlap techniques; LoongTrain~\cite{wang2024loongtrain} employs 2D parallelism across sequence and head dimensions to support million-scale token training; DistFlashAttn~\cite{fang2024distflashattn} extends FlashAttention to distributed scenarios; Colossal-AI~\cite{bian2023colossal} and LightSeq~\cite{li2023lightseq} provide unified distributed training frameworks. These works focus on the combination and scheduling of parallelism strategies, while HCMS focuses on the underlying communication-computation overlap implementation based on Ulysses; the two are orthogonal and composable.

In summary, existing sequence parallelism methods have the following limitations in communication-computation overlap: (1) Ring Attention's overlap granularity is constrained by block size, and requires $P-1$ communication rounds; (2) Ulysses achieves fewer communication rounds but lacks overlap mechanisms in its original implementation; (3) Tile-based overlap methods require modifications to underlying attention kernels, limiting their compatibility. HCMS addresses these gaps by introducing head-level chunking that enables fine-grained overlap while maintaining native compatibility with existing attention implementations.

\subsection{Communication-Computation Overlap}

Communication-computation overlap is a classical technique in distributed optimization~\cite{jangda2022breaking,jangda2021coconet}, widely applied in gradient synchronization~\cite{sergeev2018horovod,zhao2023pytorch}, ZeRO~\cite{rajbhandari2020zero} parameter prefetching, and pipeline parallelism~\cite{huang2019gpipe,narayanan2019pipedream,qi2024zero}. In the domain of distributed attention, tile-based methods~\cite{overlapattn2025,zhang2024overlap} achieve communication-computation overlap at the matrix level but require modifications to underlying attention kernels. In contrast, HCMS operates at a higher abstraction level---the attention head dimension---and is natively compatible with existing attention optimizations such as FlashAttention and SDPA, enabling orthogonal composition.

% ============================================================================
% 3. Method
% ============================================================================
\section{Method}

\subsection{Problem Formulation}

Consider all-to-all based sequence parallel attention~\cite{jacobs2023deepspeedulyssesoptimizationsenabling}. Input $\bm{X} \in \mathbb{R}^{B \times (L/P) \times D}$ is distributed across $P$ GPUs, with the forward pass consisting of: QKV projection $\rightarrow$ RoPE~\cite{su2024roformer} positional encoding $\rightarrow$ input all-to-all $\rightarrow$ attention computation $\rightarrow$ output all-to-all $\rightarrow$ output projection. The input all-to-all rearranges data from sequence-sharded to head-sharded, and the output all-to-all performs the inverse transformation. In the baseline implementation, communication and computation execute strictly in serial, with total execution time:
\begin{equation}
    T_{baseline} = T_{comm} + T_{attn} + T_{other}, \quad T_{comm} = T_{in} + T_{out}
\end{equation}
where $T_{other}$ represents fixed overhead such as QKV projection and positional encoding. Defining the communication ratio $\rho = T_{comm}/T_{baseline}$, under 4-8 GPU PCIe configurations $\rho$ typically ranges from 15\% to 40\%, making communication-computation overlap a substantial optimization opportunity.

\subsection{Method Overview}

\begin{figure*}[t]
\centering
\vspace{-0.5em}
\includegraphics[width=0.82\textwidth]{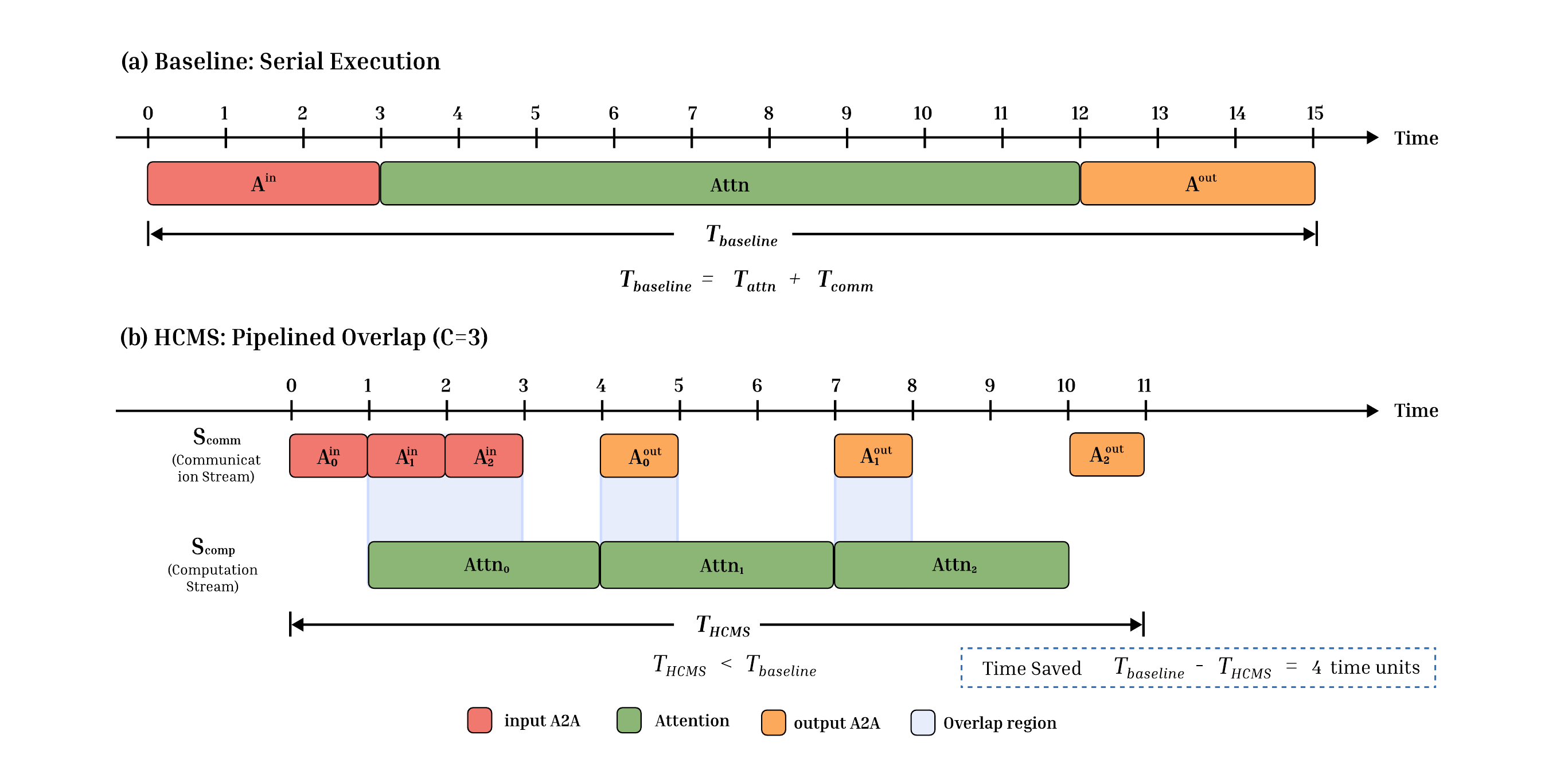}
\vspace{-1em}
\caption{HCMS method overview. (a) Baseline executes communication and computation serially; (b) HCMS decomposes operations into $C$ chunks, with dual streams executing in parallel to achieve communication-computation overlap.}
\label{fig:hcms_overview}
\vspace{-0.5em}
\end{figure*}

The core idea of HCMS is to partition attention heads into multiple chunks and achieve pipelined overlap between communication and computation through dual CUDA streams. Figure~\ref{fig:hcms_overview} illustrates the overall architecture.

As shown in the figure, HCMS comprises three key components: (1) Head Chunking partitions attention heads into $C$ independent chunks; (2) the dual-stream pipeline uses communication stream $S_{comm}$ and compute stream $S_{comp}$ for overlapped execution; (3) CUDA Event synchronization ensures correctness of data dependencies. We detail each component below.

\subsection{Head Chunking Strategy}

\begin{property}[Computational Independence of Multi-Head Attention]
\label{prop:independence}
In multi-head attention, given a total of $H$ attention heads, for any two distinct heads $h_i$ and $h_j$ ($i \neq j$):
\begin{equation}
    \bm{O}_{h_i} = f(\bm{Q}_{h_i}, \bm{K}_{h_i}, \bm{V}_{h_i})
\end{equation}
The output of head $h_i$ depends only on head $h_i$'s own inputs and is independent of other heads.
\end{property}

Based on Property~\ref{prop:independence}, we can partition $H$ attention heads into $C$ chunks, with each chunk independently completing communication and computation. Let the $c$-th chunk contain $H_c$ heads:
\begin{equation}
    H = \sum_{c=0}^{C-1} H_c, \quad H_c = \left\lfloor \frac{H}{C} \right\rfloor + \mathbb{1}_{c < H \bmod C}
\end{equation}

After partitioning, the QKV tensors are correspondingly decomposed:
\begin{equation}
    \bm{Q} = [\bm{Q}_0 \| \bm{Q}_1 \| \cdots \| \bm{Q}_{C-1}]
\end{equation}
where $\|$ denotes concatenation along the head dimension, and $\bm{Q}_c \in \mathbb{R}^{B \times (L/P) \times H_c \times D}$.

\begin{lemma}[No Data Dependency Between Chunks]
\label{lemma:no_dep}
For any two chunks $i$ and $j$ ($i \neq j$), the computation of chunk $i$ does not depend on the communication results of chunk $j$:
\begin{equation}
    \bm{O}_i = g(\text{AllToAll}(\bm{Q}_i, \bm{K}_i, \bm{V}_i))
\end{equation}
which is independent of $\bm{Q}_j, \bm{K}_j, \bm{V}_j$ and their communication results.
\end{lemma}

Lemma~\ref{lemma:no_dep} indicates that communication and computation of different chunks can be scheduled independently, providing the theoretical foundation for pipelined overlap.

\subsection{Dual-Stream Pipeline Architecture}

Based on Head Chunking, we design a dual CUDA stream pipeline architecture:

\begin{itemize}
    \item Communication stream $S_{comm}$: dedicated to executing all-to-all collective communication
    \item Compute stream $S_{comp}$: dedicated to executing attention computation
\end{itemize}

\begin{figure*}[t]
\centering
\vspace{-0.5em}
\includegraphics[width=0.85\textwidth]{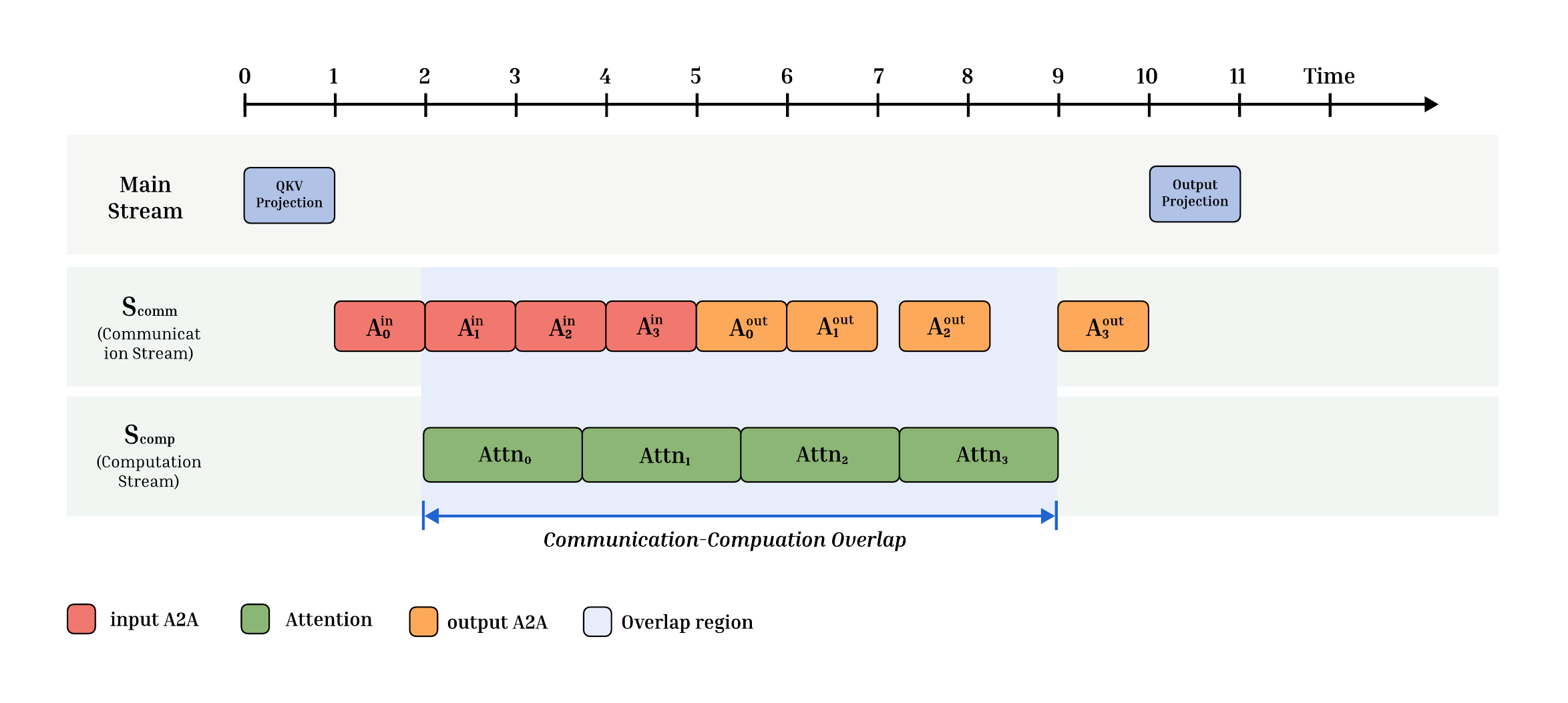}
\vspace{-1em}
\caption{HCMS pipeline execution timeline. Communication stream $S_{comm}$ and compute stream $S_{comp}$ execute in parallel, with dashed lines indicating CUDA Event synchronization dependencies.}
\label{fig:pipeline}
\vspace{-0.5em}
\end{figure*}

Pipeline execution consists of three phases, as shown in Figure~\ref{fig:pipeline}:

Phase 1 - Input Communication: The communication stream sequentially initiates input all-to-all for each chunk, recording event $E_c^{comm}$ upon completion of each chunk's communication.

Phase 2 - Attention Computation: The compute stream monitors $E_c^{comm}$ events; once chunk $c$'s input data is ready, computation begins immediately. Event $E_c^{comp}$ is recorded upon completion.

Phase 3 - Output Communication: The communication stream monitors $E_c^{comp}$ events; once chunk $c$'s computation completes, output all-to-all is initiated immediately.

\subsection{CUDA Event Synchronization Mechanism}

Pipeline correctness depends on precise dependency management. We define two types of CUDA Events:

\begin{itemize}
    \item $E_c^{comm}$: recorded when chunk $c$'s input all-to-all completes
    \item $E_c^{comp}$: recorded when chunk $c$'s attention computation completes
\end{itemize}

Dependencies are expressed as:
\begin{align}
    \text{Attn}_c &\leftarrow \text{wait}(E_c^{comm}) \\
    A_c^{out} &\leftarrow \text{wait}(E_c^{comp})
\end{align}

Key observation: Due to Lemma~\ref{lemma:no_dep}, $A_{c+1}^{in}$ does not need to wait for $\text{Attn}_c$ or $A_c^{out}$ to complete. This means:
\begin{equation}
    A_{c+1}^{in} \parallel \text{Attn}_c \parallel A_c^{out}
\end{equation}
All three can execute in parallel, subject to hardware resource constraints.

\subsection{Algorithm Description}

Algorithm~\ref{alg:hcms} provides the complete pseudocode for HCMS.

\begin{algorithm}[htbp]
\caption{Head-Chunked Multi-Stream Pipeline (HCMS)}
\label{alg:hcms}
\begin{algorithmic}[1]
\REQUIRE Input $\bm{X}$, num\_chunks $C$
\ENSURE Output $\bm{Y}$

\STATE // Main Stream
\STATE $\bm{Q}, \bm{K}, \bm{V} \leftarrow$ QKVProjection($\bm{X}$)
\STATE $\bm{Q}, \bm{K} \leftarrow$ ApplyRoPE($\bm{Q}, \bm{K}$)
\STATE $\{\bm{Q}_c, \bm{K}_c, \bm{V}_c\}_{c=0}^{C-1} \leftarrow$ SplitByHeads($C$)
\STATE Record $E_{main}$

\STATE // Communication Stream: Input A2A
\STATE $S_{comm}$.wait($E_{main}$)
\FOR{$c = 0$ to $C-1$}
    \STATE $\bm{Q}'_c, \bm{K}'_c, \bm{V}'_c \leftarrow$ AllToAll($\bm{Q}_c, \bm{K}_c, \bm{V}_c$)
    \STATE Record $E_c^{comm}$ on $S_{comm}$
\ENDFOR

\STATE // Compute Stream: Attention
\STATE $S_{comp}$.wait($E_{main}$)
\FOR{$c = 0$ to $C-1$}
    \STATE $S_{comp}$.wait($E_c^{comm}$)
    \STATE $\bm{O}'_c \leftarrow$ Attention($\bm{Q}'_c, \bm{K}'_c, \bm{V}'_c$) \COMMENT{FlashAttention or SDPA}
    \STATE Record $E_c^{comp}$ on $S_{comp}$
\ENDFOR

\STATE // Communication Stream: Output A2A
\FOR{$c = 0$ to $C-1$}
    \STATE $S_{comm}$.wait($E_c^{comp}$)
    \STATE $\bm{O}_c \leftarrow$ AllToAll($\bm{O}'_c$)
\ENDFOR

\STATE // Main Stream: Merge \& Project
\STATE Synchronize($S_{comm}$, $S_{comp}$)
\STATE $\bm{O} \leftarrow$ Concat($\{\bm{O}_c\}_{c=0}^{C-1}$)
\STATE $\bm{Y} \leftarrow$ OutputProjection($\bm{O}$)
\RETURN $\bm{Y}$
\end{algorithmic}
\end{algorithm}

\subsection{Theoretical Analysis}
\label{sec:theory}

We establish a theoretical performance model for HCMS, deriving the optimal chunk count and speedup upper bound; detailed derivations are provided in Appendix~\ref{appendix:theory}. Let communication time be $T_{comm}$, per-chunk overhead be $\beta$, and communication ratio $\rho = T_{comm}/T_{total}$. The core conclusions are:

\begin{theorem}[Optimal Chunk Count]
\label{thm:optimal_chunk}
The chunk count that minimizes HCMS execution time is $C^* = \sqrt{T_{comm}/\beta}$.
\end{theorem}

\begin{theorem}[Speedup Upper Bound]
\label{thm:upper_bound}
When $\beta \to 0$ and $C \to \infty$, the theoretical upper bound of HCMS speedup is $S_{max} = 1/(1-\rho)$.
\end{theorem}

Applicability: HCMS benefits are positively correlated with communication ratio $\rho$. When $\rho > 20\%$, HCMS is recommended and can achieve 3\%-24\% speedup; when $\rho < 10\%$, benefits are limited to less than 1\%. Factors affecting $\rho$ include GPU count, sequence length, and interconnect type, with PCIe interconnect scenarios typically having higher $\rho$ than NVLink, making HCMS benefits more pronounced.

\subsection{Comparison with Ring Attention and Ulysses}

Table~\ref{tab:comparison} compares HCMS with existing methods across multiple dimensions.

\begin{table}[htbp]
\centering
\caption{Comparison of HCMS with existing methods}
\label{tab:comparison}
\scriptsize
\setlength{\tabcolsep}{3pt}
\begin{tabular}{lccc}
\toprule
Property & Ring Attn & Ulysses & HCMS \\
\midrule
Comm. Pattern & P2P Ring & All-to-All & All-to-All \\
Comm. Rounds & $P-1$ & 2 & 2 \\
Comm. Volume & $O(LD)$ & $O(LD)$ & $O(LD)$ \\
\midrule
Overlap & Block-level & None & Chunk-level \\
Granularity & Block & - & Head \\
\midrule
FlashAttn & Requires mod. & Native & Native \\
Causal Mask & Native & Native & Native \\
\midrule
Optimal & Long seq. & Med-long & High $\rho$ \\
\bottomrule
\end{tabular}
\end{table}

Comparison with Ring Attention: Ring Attention's overlap occurs at the block level, communicating one complete KV block per round, while HCMS's overlap occurs at the head level with finer granularity. Regarding communication rounds, Ring Attention requires $P-1$ rounds of serial communication, while HCMS requires only 2 rounds. The two have different optimal scenarios: Ring Attention excels in very long sequence, memory-constrained scenarios, while HCMS performs better in medium-long sequence scenarios with high communication ratio.

Comparison with Ulysses: The two differ fundamentally in their optimization dimensions. Ulysses focuses on the choice between all-to-all and P2P communication patterns, treating communication and computation as atomic operations executed serially; HCMS introduces Head Chunking as a new partitioning abstraction, leveraging the computational independence of multi-head attention (Property~\ref{prop:independence}) to decompose atomic operations into independently schedulable subtasks, thereby achieving fine-grained pipelining at the head dimension. This concept is general: any all-to-all based sequence parallelism scheme can introduce communication-computation overlap through Head Chunking, with HCMS's application to Ulysses being just one instance.

% ============================================================================
% 4. Experiments
% ============================================================================
\section{Experiments}
\label{sec:exp}

This section validates the effectiveness of HCMS. The core claims are:
\begin{itemize}[itemsep=0.1pt,topsep=0pt,leftmargin=*]
    \item Performance improvement: HCMS achieves 10\%-17.5\% speedup over Ulysses and 5\%-14.5\% over Ring Attention for typical video generation sequence lengths of 31K-56K tokens, validated across four GPU platforms at 2-8 GPU scales.
    \item End-to-end benefits: In the Wan2.2 video generation model, attention layers achieve 18.3\% speedup and the complete pipeline achieves 6.8\% speedup.
    \item No accuracy loss: HCMS output is identical to baseline across all configurations, supports training scenarios with 3.5\% speedup, and reduces peak memory by 8.7\%.
\end{itemize}
Appendix~\ref{appendix:exp} contains complete experimental data and supplementary analysis.

\subsection{Experimental Setup}

Hardware environment: To validate generalization, we conduct experiments on four GPU platforms with configurations shown in Table~\ref{tab:hardware}. Software environment is PyTorch~\cite{paszke2019pytorch} 2.9.1, CUDA 12.8, collective communication via NCCL~\cite{nvidia2015nccl}, with BF16~\cite{micikevicius2018mixed} numerical precision.

\begin{table}[htbp]
\centering
\caption{Experimental platform configurations}
\label{tab:hardware}
\scriptsize
\setlength{\tabcolsep}{3pt}
\begin{tabular}{lcccc}
\toprule
Platform & GPU & Mem & Interconn. & Attention \\
\midrule
A & 4$\times$L20 & 46GB & PCIe 4.0 & FlashAttn-2 \\
B & 4$\times$4090 & 24GB & PCIe 4.0 & SDPA \\
C & 4$\times$A10 & 22GB & PCIe 4.0 & SDPA \\
D & 8$\times$5090 & 32GB & PCIe 5.0 & SDPA \\
\bottomrule
\end{tabular}
\end{table}

Model configuration: We use typical parameters from video generation models~\cite{dhariwal2021diffusion,rombach2022high}, shown in Table~\ref{tab:model_config}.

\begin{table}[htbp]
\centering
\caption{Model configuration}
\label{tab:model_config}
\scriptsize
\begin{tabular}{lc}
\toprule
Parameter & Value \\
\midrule
Hidden dim. $D_{model}$ & 5120 \\
Attention heads $H$ & 40 \\
Head dimension $D$ & 128 \\
Latent resolution & $60 \times 104$ \\
Frames & 9 / 21 / 33 / 45 \\
Sequence length & 56K / 131K / 206K / 281K \\
\bottomrule
\end{tabular}
\end{table}

Evaluation metrics: We use end-to-end latency, throughput, and speedup ratio for performance evaluation, while verifying numerical consistency through maximum/mean error.

Baseline method: We use the standard implementation of DeepSpeed Ulysses~\cite{jacobs2023deepspeedulyssesoptimizationsenabling} as the baseline. Ulysses is the mainstream sequence parallelism approach, with communication pattern consisting of input all-to-all, attention computation, and output all-to-all in three stages. In the original Ulysses implementation, communication and computation execute strictly in serial, which is precisely the optimization target of HCMS.

Implementation details: HCMS employs device-level stream caching to avoid frequent creation overhead, and KV Cache uses a chunk-organized list structure to support chunked computation. When $H$ is not divisible by $C$, a ``more-first-less-later'' uneven partitioning strategy is used; experiments show performance variance is less than 1\%.

\subsection{Main Results}

Table~\ref{tab:gpu_scaling} shows performance across different GPU configurations on each platform.

\begin{table}[htbp]
\centering
\caption{Performance across different GPU counts, 131K tokens, $C=4$}
\label{tab:gpu_scaling}
\scriptsize
\begin{tabular}{lccccc}
\toprule
Platform & GPUs & Ulysses & HCMS & Speedup \\
\midrule
\multirow{2}{*}{L20+FA} & 2 & 1896.8ms & 1828.6ms & 1.037$\times$ \\
 & 4 & 996.2ms & 945.3ms & 1.054$\times$ \\
\midrule
\multirow{2}{*}{A10+SDPA} & 2 & 3424.1ms & 3007.0ms & 1.139$\times$ \\
 & 4 & 1643.2ms & 1458.8ms & 1.126$\times$ \\
\midrule
\multirow{3}{*}{5090+SDPA} & 2 & 988.4ms & 981.5ms & 1.007$\times$ \\
 & 4 & 502.3ms & 500.5ms & 1.004$\times$ \\
 & 8 & 288.3ms & 276.1ms & 1.044$\times$ \\
\bottomrule
\end{tabular}
\end{table}

From the GPU scaling perspective, speedup correlates positively with communication ratio $\rho$: the A10 platform with higher $\rho$ achieves 12.6\%-13.9\% speedup; the L20 platform achieves 3.7\%-5.4\% speedup; the RTX 5090 8-GPU configuration achieves 4.4\% speedup, validating HCMS effectiveness in large-scale parallel scenarios.

Table~\ref{tab:seq_scaling} shows performance comparison across different sequence lengths.

\begin{table}[htbp]
\centering
\caption{Performance comparison across different sequence lengths, $C=4$}
\label{tab:seq_scaling}
\scriptsize
\setlength{\tabcolsep}{3pt}
\begin{tabular}{lcccccc}
\toprule
Seq. & \multicolumn{2}{c}{L20+FA 4-GPU} & \multicolumn{2}{c}{4090+SDPA 4-GPU} & \multicolumn{2}{c}{5090 8-GPU} \\
\cmidrule(lr){2-3} \cmidrule(lr){4-5} \cmidrule(lr){6-7}
 & Ulysses & Spdup & Ulysses & Spdup & Ulysses & Spdup \\
\midrule
56K & 234.5ms & 1.10$\times$ & 182.8ms & 1.18$\times$ & 73.3ms & 1.16$\times$ \\
131K & 996.4ms & 1.06$\times$ & 725.9ms & 1.07$\times$ & 288.3ms & 1.05$\times$ \\
206K & 2262.2ms & 1.04$\times$ & 1617.3ms & 1.06$\times$ & 631.0ms & 1.05$\times$ \\
281K & 4046.3ms & 1.03$\times$ & 2868.1ms & 1.04$\times$ & 1112.9ms & 1.01$\times$ \\
\bottomrule
\end{tabular}
\end{table}

Experimental results show clear sequence length dependency: shorter sequences yield higher speedups. The 56K sequence achieves 17.5\% speedup on 4090 4-GPU configuration and 16.4\% speedup on 5090 8-GPU configuration. The 4090 platform achieves overall higher speedups than L20 because both platforms use PCIe 4.0 interconnect with similar communication time, while 4090 has shorter total execution time (182.8ms vs 234.5ms), resulting in higher communication ratio $\rho$ and thus greater HCMS benefits. The A10 platform achieves 12.6\%-13.9\% speedup at 131K tokens, further validating the ``higher communication ratio, greater benefits'' principle. For 8-GPU scalability, the 5090 platform still achieves 16.4\% speedup at 56K tokens, while benefits decrease to 0.9\% at 281K tokens, consistent with theoretical expectations. All four platforms (L20, 4090, A10, 5090) at 2-8 GPU scales validate HCMS effectiveness, demonstrating cross-hardware, cross-scale generalization.

Reason analysis: Attention computation complexity is $O(L^2)$, while communication complexity is $O(L)$. As sequences grow longer, computation time increases much faster than communication time, causing communication ratio $\rho$ to decrease and HCMS overlap benefits to diminish.

\begin{figure}[htbp]
\centering
\begin{tikzpicture}
\begin{axis}[
    width=0.9\columnwidth,
    height=5cm,
    xlabel={Sequence Length (K tokens)},
    ylabel={Speedup},
    xmin=40, xmax=300,
    ymin=1.0, ymax=1.20,
    xtick={56, 131, 206, 281},
    ytick={1.0, 1.04, 1.08, 1.12, 1.175},
    grid=major,
    legend pos=north east,
    mark size=2pt,
]
\addplot[color=blue, mark=*, thick] coordinates {
    (56, 1.103)
    (131, 1.055)
    (206, 1.037)
    (281, 1.030)
};
\addlegendentry{L20 4-GPU}

\addplot[color=red, mark=square*, thick] coordinates {
    (56, 1.175)
    (131, 1.071)
    (206, 1.060)
    (281, 1.041)
};
\addlegendentry{4090 4-GPU}

\addplot[color=green!60!black, mark=triangle*, thick] coordinates {
    (56, 1.164)
    (131, 1.049)
    (206, 1.046)
    (281, 1.009)
};
\addlegendentry{5090 8-GPU}

\addplot[color=gray, dashed, domain=40:300] {1.0};
\addlegendentry{Baseline}
\end{axis}
\end{tikzpicture}
\caption{Speedup versus sequence length. All three configurations show greater HCMS benefits for shorter sequences. The 5090 8-GPU configuration validates effectiveness in large-scale parallel scenarios.}
\label{fig:speedup_vs_seq}
\end{figure}
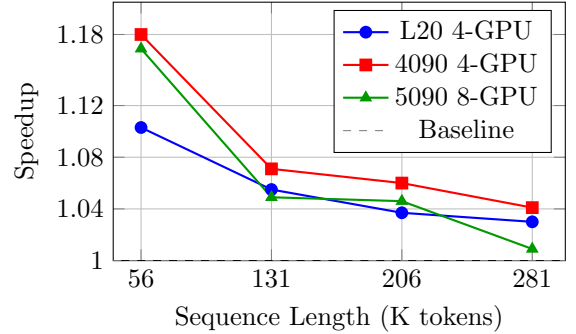

Figure~\ref{fig:speedup_vs_seq} visualizes this trend.

Comparison with Ring Attention: Table~\ref{tab:ring_comparison} presents a three-way comparison of HCMS, Ring Attention, and Ulysses. For typical video generation sequence lengths of 31K-56K tokens, HCMS leads comprehensively: 14.5\% faster than Ring at 31K tokens, 5.0\% faster at 56K. For longer sequences above 131K, Ring is slightly better by 1.7\%, with the crossover point at approximately 60-80K tokens.

\begin{table}[htbp]
\centering
\caption{Three-way comparison: HCMS vs Ring Attention vs Ulysses, L20+FA 4-GPU}
\label{tab:ring_comparison}
\scriptsize
\begin{tabular}{lcccc}
\toprule
Seq. & Ulysses & Ring & HCMS & Best \\
\midrule
31K & 95.5ms & 95.4ms & 83.3ms & HCMS (+14.5\%) \\
56K & 234.7ms & 223.3ms & 212.7ms & HCMS (+5.0\%) \\
131K & 996.9ms & 929.4ms & 945.4ms & Ring (+1.7\%) \\
\bottomrule
\end{tabular}
\end{table}

\subsection{Ablation Studies}

Table~\ref{tab:chunk_ablation} shows the effect of different chunk counts on performance, with experimental configuration of 4 GPUs and 131K tokens.

\begin{table}[htbp]
\centering
\caption{Performance with different chunk counts}
\label{tab:chunk_ablation}
\scriptsize
\begin{tabular}{lccc}
\toprule
Chunks & Latency (ms) & Speedup & Rel. to Best \\
\midrule
1 (Ulysses) & 996.2 & 1.000$\times$ & -5.2\% \\
2 & 951.5 & 1.047$\times$ & -0.7\% \\
4 & 945.3 & 1.054$\times$ & -0.1\% \\
5 & 944.5 & 1.055$\times$ & Best \\
8 & 963.8 & 1.034$\times$ & -2.0\% \\
10 & 970.9 & 1.026$\times$ & -2.7\% \\
\bottomrule
\end{tabular}
\end{table}

Results show that as $C$ increases from 1 to 5, speedup continuously improves because more chunks enable finer-grained communication-computation overlap. However, performance begins to decline when $C$ exceeds 5, due to accumulated Event synchronization overhead and efficiency loss from overly fine computation granularity. In this configuration, the optimal point is $C=5$, but $C=4$ is already near-optimal and simpler to implement.

Uneven partitioning robustness: When $H$ is not divisible by $C$, a ``more-first-less-later'' strategy is used. Table~\ref{tab:uneven_chunking} shows partitioning schemes and performance for different $C$ values, with configuration $H=40$, 4 GPUs, $H_{per\_rank}=10$.

\begin{table}[htbp]
\centering
\caption{Uneven chunk partitioning performance}
\label{tab:uneven_chunking}
\scriptsize
\setlength{\tabcolsep}{4pt}
\begin{tabular}{ccccc}
\toprule
$C$ & Type & Chunk Sizes & Imbal. & Speedup \\
\midrule
2 & Even & [5, 5] & 0\% & 1.049$\times$ \\
3 & Uneven & [4, 3, 3] & 30\% & 1.056$\times$ \\
4 & Uneven & [3, 3, 2, 2] & 40\% & 1.055$\times$ \\
5 & Even & [2, 2, 2, 2, 2] & 0\% & 1.055$\times$ \\
6 & Uneven & [2, 2, 2, 2, 1, 1] & 60\% & 1.052$\times$ \\
\bottomrule
\end{tabular}
\end{table}

Even partitioning achieves average speedup of 1.044$\times$, while uneven partitioning achieves 1.048$\times$, with only 0.4\% difference. This indicates that load imbalance effects are masked by pipelined execution, and the optimal $C$ value need not satisfy divisibility.

Correctness and training support: HCMS output is identical to baseline across all configurations, with maximum error of 0. HCMS is compatible with PyTorch autograd, achieving 3.5\% speedup for forward+backward combined execution and 8.7\% reduction in peak memory; see Appendix~\ref{appendix:exp} for details.

\subsection{End-to-End Video Generation Validation}

To validate HCMS effectiveness in real applications, we integrate it into the Wan2.2 video generation model~\cite{wang2024opensora} for end-to-end experiments.

Experimental setup: We use the Wan2.2 TI2V-5B model (Text-Image-to-Video 5B parameters) with 24 attention heads, 128 head dimension, 30-layer DiT~\cite{peebles2023scalable}, resolution 1280$\times$704, and 4 GPUs with Ulysses sequence parallelism.

Attention layer benchmark: Table~\ref{tab:wan22_attn} shows attention layer performance for the TI2V-5B configuration with 24 heads and 4 GPUs across different sequence lengths.

\begin{table}[htbp]
\centering
\caption{Wan2.2 TI2V-5B attention layer performance, 24 heads, 4 GPUs, $C=6$}
\label{tab:wan22_attn}
\scriptsize
\begin{tabular}{lcccc}
\toprule
Frames & Seq. Length & Ulysses & HCMS & Speedup \\
\midrule
17 frames & 17.6K & 16.35ms & 13.36ms & +18.3\% \\
49 frames & 43.2K & 72.61ms & 61.87ms & +14.8\% \\
81 frames & 70.4K & 167.56ms & 152.91ms & +8.7\% \\
\bottomrule
\end{tabular}
\end{table}

End-to-end performance: Table~\ref{tab:wan22_e2e} shows end-to-end performance for the complete video generation pipeline.

\begin{table}[htbp]
\centering
\caption{Wan2.2 TI2V-5B end-to-end performance, 4 GPUs, $C=6$}
\label{tab:wan22_e2e}
\scriptsize
\begin{tabular}{lccccc}
\toprule
Frames & Video Duration & Ulysses & HCMS & Speedup \\
\midrule
17 frames & 0.7s & 1.69s/it & 1.67s/it & +1.2\% \\
49 frames & 2.0s & 2.34s/it & 2.18s/it & +6.8\% \\
\bottomrule
\end{tabular}
\end{table}

For attention layer performance, the 17-frame short video configuration achieves 18.3\% speedup, and the 49-frame medium-length video configuration achieves 14.8\% speedup. For end-to-end performance, 49-frame video per-step inference time decreases from 2.34s to 2.18s, achieving measured speedup of 6.8\%. In this configuration, the optimal chunk count is $C=6$, i.e., 24 heads partitioned into 6 chunks with 4 heads per chunk, where more chunks provide better pipeline overlap.

End-to-end benefit analysis: Attention layers achieve 14.8\% speedup, but end-to-end is only 6.8\%, because the video generation pipeline includes other components such as T5 text encoding and VAE decoding. For pure DiT inference scenarios, HCMS end-to-end benefits will be closer to attention layer speedup.

% ============================================================================
% 5. Conclusion
% ============================================================================
\section{Conclusion}

This paper proposes Head-Chunked Multi-Stream Pipeline (HCMS), optimizing communication-computation overlap for sequence parallel attention in medium-long sequence scenarios of 31K-100K tokens such as video generation. By partitioning attention heads into independent chunks through the Head Chunking strategy and achieving fine-grained pipelining via dual CUDA streams, efficient parallelization of communication and computation is achieved without changing computational semantics. Theoretical analysis shows HCMS benefits are positively correlated with communication ratio $\rho$, and its use is recommended when $\rho > 20\%$.

Experiments validate effectiveness across four GPU platforms at 2-8 GPU scales: for typical video generation sequence lengths of 31K-56K tokens, HCMS achieves 10\%-17.5\% speedup over the Ulysses baseline and 5\%-14.5\% over Ring Attention; in Wan2.2 end-to-end validation, attention layers achieve 18.3\% speedup and the complete pipeline achieves 6.8\% speedup. HCMS supports uneven partitioning with performance variance less than 1\%, is compatible with PyTorch autograd with 3.5\% training speedup, and is orthogonally composable with frameworks such as USP and LoongTrain.

\section{Discussion}

Limitations: HCMS benefits are limited in compute-dominated scenarios where $\rho < 10\%$, such as 281K token long sequences achieving only 3.0\% speedup. Additionally, when chunk count $C$ is too large, Event synchronization overhead accumulates noticeably, and implementation introduces multi-stream management complexity.

Future work: This paper validates up to 8-GPU scenarios; future work will extend to 16-GPU and cross-node scenarios. Other directions include: applying HCMS to backward pass gradient communication overlap, and implementing adaptive scheduling mechanisms that dynamically select $C$ based on runtime profiling.

\bibliographystyle{plain}
\bibliography{references}

% ============================================================================
% Appendix
% ============================================================================
\appendix

\section{Detailed Theoretical Derivations}
\label{appendix:theory}

This section provides complete derivations for the HCMS theoretical performance model.

\subsection{Notation}

For convenience of analysis, we define the notation shown in Table~\ref{tab:notation}:

\begin{table}[htbp]
\centering
\caption{Notation for theoretical analysis}
\label{tab:notation}
\scriptsize
\setlength{\tabcolsep}{3pt}
\begin{tabular}{cl}
\toprule
Symbol & Meaning \\
\midrule
$T_{comm}$ & Total comm. time, $T_{comm} = T_{a2a,in} + T_{a2a,out}$ \\
$T_{attn}$ & Total attention computation time \\
$T_0$ & Fixed overhead (QKV proj., output proj.) \\
$C$ & Number of chunks \\
$\beta$ & Per-chunk overhead (Event sync, kernel launch) \\
$\rho$ & Communication ratio, $\rho = T_{comm} / T_{total}$ \\
\bottomrule
\end{tabular}
\end{table}

\subsection{HCMS Time Model}

Baseline execution time. In serial execution mode, total time is the sum of all stages:
\begin{equation}
    T_{base} = T_0 + T_{comm} + T_{attn}
\end{equation}

HCMS execution time. In pipeline mode, $H$ attention heads are partitioned into $C$ chunks. Due to no data dependency between chunks (Lemma~\ref{lemma:no_dep}), computation of chunk $i$ can execute in parallel with communication of chunk $i+1$. Let per-chunk overhead be $\beta$, then:
\begin{equation}
\label{eq:hcms_time}
    T_{hcms}(C) = T_0 + \frac{T_{comm}}{C} + T_{attn} + C \cdot \beta
\end{equation}

\subsection{Optimal Chunk Count Derivation}

Taking the derivative of Equation~(\ref{eq:hcms_time}) with respect to $C$ and setting it to zero yields $C^* = \sqrt{T_{comm}/\beta}$. The second-order condition confirms this is a minimum.

\subsection{Speedup Analysis}

Let communication ratio $\rho = T_{comm}/T_{total}$, the speedup is:
\begin{equation}
\label{eq:speedup}
    S(C) = \frac{1}{1 - \rho \cdot \frac{C-1}{C} + \frac{C \cdot \beta}{T_{total}}}
\end{equation}

When $\beta \to 0$ and $C \to \infty$, the speedup upper bound is $S_{max} = 1/(1-\rho)$.

\subsection{Theoretical Predictions vs. Experimental Validation}

Table~\ref{tab:theory_validation} shows comparison of theoretical predictions with measurements on the RTX 5090 platform. Time model prediction errors are all less than 1.4\%.

\begin{table}[htbp]
\centering
\caption{Theoretical predictions vs. measurements, RTX 5090 8-GPU}
\label{tab:theory_validation}
\scriptsize
\setlength{\tabcolsep}{3pt}
\begin{tabular}{@{}cc@{\hspace{4pt}}c@{\hspace{4pt}}c@{\hspace{6pt}}cc@{\hspace{4pt}}c@{\hspace{4pt}}c@{}}
\toprule
$H$ & Seq & $\rho$ & $\beta$ & \multicolumn{2}{c}{$C^*$} & Error & Speedup \\
\cmidrule(lr){5-6}
 & & & (ms) & Pred. & Meas. & & \\
\midrule
40 & 56K & 39.6\% & 1.41 & 5 & 5 & 0.58\% & 24.0\% \\
40 & 131K & 22.1\% & 6.83 & 1 & 5 & 0.52\% & 4.2\% \\
24 & 56K & 38.1\% & 1.23 & 3 & 3 & 1.38\% & 16.0\% \\
24 & 131K & 21.7\% & 5.52 & 3 & 3 & 1.15\% & 2.9\% \\
\bottomrule
\end{tabular}
\end{table}

\subsection{Applicability Guidelines}

Table~\ref{tab:rho_analysis} summarizes applicability for different communication ratios $\rho$.

\begin{table}[htbp]
\centering
\caption{Relationship between communication ratio and speedup}
\label{tab:rho_analysis}
\scriptsize
\setlength{\tabcolsep}{3pt}
\begin{tabular}{cccc}
\toprule
Comm. Ratio $\rho$ & Theor. $S_{max}$ & Meas. Spdup & Recomm. \\
\midrule
$>35\%$ & $>1.54\times$ & 16\%-24\% & Highly rec. \\
20\%-35\% & 1.25$\times$-1.54$\times$ & 3\%-6\% & Recommended \\
10\%-20\% & 1.11$\times$-1.25$\times$ & 1\%-3\% & Optional \\
$<10\%$ & $<1.11\times$ & $<1\%$ & Limited \\
\bottomrule
\end{tabular}
\end{table}

\section{System Architecture}

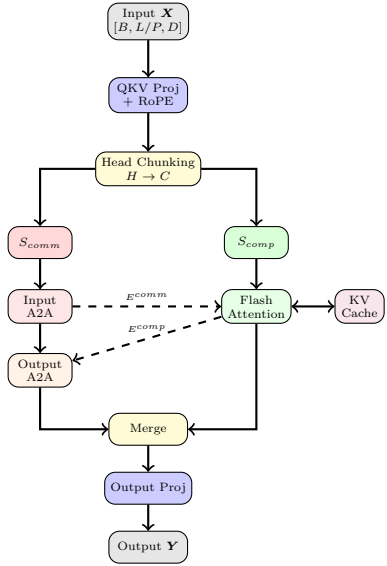
\begin{figure}[htbp]
\centering
\begin{tikzpicture}[
    scale=0.65,
    transform shape,
    box/.style={rectangle, draw, rounded corners, minimum height=0.65cm, minimum width=1.6cm, font=\scriptsize, align=center},
    arrow/.style={->, thick},
    dasharrow/.style={->, dashed, thick}
]
\node[box, fill=gray!20] (input) at (0, 5) {Input $\bm{X}$\\$[B, L/P, D]$};
\node[box, fill=blue!20] (qkv) at (0, 3.5) {QKV Proj\\+ RoPE};
\node[box, fill=yellow!20] (chunk) at (0, 2) {Head Chunking\\$H \rightarrow C$};
\node[box, fill=red!15, minimum width=1.3cm] (scomm) at (-2.2, 0.5) {$S_{comm}$};
\node[box, fill=green!15, minimum width=1.3cm] (scomp) at (2.2, 0.5) {$S_{comp}$};
\node[box, fill=red!10, minimum width=1.3cm] (a2a_in) at (-2.2, -0.8) {Input\\A2A};
\node[box, fill=green!10, minimum width=1.3cm] (attn) at (2.2, -0.8) {Flash\\Attention};
\node[box, fill=orange!10, minimum width=1.3cm] (a2a_out) at (-2.2, -2.1) {Output\\A2A};
\node[box, fill=yellow!20] (merge) at (0, -3.3) {Merge};
\node[box, fill=blue!20] (proj) at (0, -4.5) {Output Proj};
\node[box, fill=gray!20] (output) at (0, -5.7) {Output $\bm{Y}$};
\node[box, fill=purple!10, minimum width=1.0cm] (kv) at (4.3, -0.8) {KV\\Cache};
\draw[arrow] (input) -- (qkv);
\draw[arrow] (qkv) -- (chunk);
\draw[arrow] (chunk) -| (scomm);
\draw[arrow] (chunk) -| (scomp);
\draw[arrow] (scomm) -- (a2a_in);
\draw[arrow] (scomp) -- (attn);
\draw[arrow] (a2a_in) -- (a2a_out);
\draw[dasharrow] (a2a_in) -- node[above, font=\tiny] {$E^{comm}$} (attn);
\draw[dasharrow] (attn) -- node[above, font=\tiny] {$E^{comp}$} (a2a_out);
\draw[arrow] (a2a_out) |- (merge);
\draw[arrow] (attn) |- (merge);
\draw[arrow] (merge) -- (proj);
\draw[arrow] (proj) -- (output);
\draw[<->, thick] (attn) -- (kv);
\end{tikzpicture}
\caption{HCMS system architecture}
\label{fig:architecture}
\end{figure}

Figure~\ref{fig:architecture} shows the complete system architecture of HCMS.

\section{Detailed Experimental Data}
\label{appendix:exp}

\subsection{Correctness and Training Validation}

HCMS output is identical to baseline across all configurations, as shown in Table~\ref{tab:correctness}. Forward+backward combined execution achieves 3.5\% speedup (Table~\ref{tab:backward}), with 8.7\% reduction in peak memory (Table~\ref{tab:memory}).

\begin{table}[htbp]
\centering
\caption{Numerical correctness validation, BF16 precision}
\label{tab:correctness}
\scriptsize
\begin{tabular}{lccc}
\toprule
GPU Rank & Max Diff & Mean Diff & Status \\
\midrule
0 & 0.0 & 0.0 & PASS \\
1 & 0.0 & 0.0 & PASS \\
2 & 0.0 & 0.0 & PASS \\
3 & 0.0 & 0.0 & PASS \\
\bottomrule
\end{tabular}
\end{table}

\begin{table}[htbp]
\centering
\caption{Forward+backward combined performance, 4 GPUs, 131K tokens}
\label{tab:backward}
\scriptsize
\begin{tabular}{lcc}
\toprule
Method & Time (ms) & Speedup \\
\midrule
Ulysses & 3292.5 & 1.000$\times$ \\
HCMS ($C$=5) & 3181.4 & 1.035$\times$ \\
\bottomrule
\end{tabular}
\end{table}

\begin{table}[htbp]
\centering
\caption{GPU peak memory comparison, 4 GPUs, 131K tokens}
\label{tab:memory}
\scriptsize
\begin{tabular}{lcc}
\toprule
Method & Peak Memory (GB) & Relative Change \\
\midrule
Ulysses & 6.767 & --- \\
HCMS & 6.178 & -8.7\% \\
\bottomrule
\end{tabular}
\end{table}

\subsection{Complete Experimental Results}

Table~\ref{tab:full_results} summarizes complete results for all experimental configurations.

\begin{table}[H]
\centering
\caption{Complete experimental results}
\label{tab:full_results}
\scriptsize
\setlength{\tabcolsep}{2.5pt}
\begin{tabular}{lcccccc}
\toprule
Platform & GPU & Seq & C & Ulysses & HCMS & Spdup \\
\midrule
\multirow{5}{*}{L20+FA} & 4 & 56K & 4 & 234.5 & 212.6 & 1.10$\times$ \\
& 4 & 131K & 4 & 996.2 & 945.3 & 1.05$\times$ \\
& 4 & 206K & 4 & 2262.2 & 2182.1 & 1.04$\times$ \\
& 4 & 281K & 4 & 4046.3 & 3928.4 & 1.03$\times$ \\
& 2 & 131K & 4 & 1896.8 & 1828.6 & 1.04$\times$ \\
\midrule
\multirow{4}{*}{4090+SDPA} & 4 & 56K & 4 & 182.8 & 155.6 & 1.18$\times$ \\
& 4 & 131K & 4 & 725.9 & 677.8 & 1.07$\times$ \\
& 4 & 206K & 4 & 1617.3 & 1525.8 & 1.06$\times$ \\
& 4 & 281K & 4 & 2868.1 & 2755.1 & 1.04$\times$ \\
\midrule
\multirow{2}{*}{A10+SDPA} & 2 & 131K & 4 & 3424.1 & 3007.0 & 1.14$\times$ \\
& 4 & 131K & 4 & 1643.2 & 1458.8 & 1.13$\times$ \\
\midrule
\multirow{6}{*}{5090+SDPA} & 2 & 131K & 4 & 988.4 & 981.5 & 1.01$\times$ \\
& 4 & 131K & 4 & 502.3 & 500.5 & 1.00$\times$ \\
& 8 & 131K & 4 & 288.3 & 276.1 & 1.04$\times$ \\
& 8 & 56K & 4 & 73.3 & 63.0 & 1.16$\times$ \\
& 8 & 206K & 4 & 631.0 & 603.3 & 1.05$\times$ \\
& 8 & 281K & 4 & 1112.9 & 1102.4 & 1.01$\times$ \\
\bottomrule
\end{tabular}
\end{table}

\end{document}